\documentclass[twocolumn,showpacs,preprintnumbers,amssymb,aps,pra]{revtex4}
\usepackage{graphicx}
\usepackage{dcolumn}
\usepackage{bm}
\usepackage{latexsym,epsfig}

\usepackage{array}
\usepackage{color}
\usepackage{amstext}

\begin{document}

\title{Relativistic coupled-cluster theory analysis of unusually large correlation effects in the determination of $g_j$ factors in Ca$^+$}
\vspace*{0.5cm}

\author{B. K. Sahoo$^a$ \footnote{Email: bijaya@prl.res.in} and Pradeep Kumar$^{a,b}$}
\affiliation{$^a$Atomic, Molecular and Optical Physics Division, Physical Research Laboratory, Navrangpura, Ahmedabad-380009, India}
\affiliation{$^b$Indian Institute of Technology Gandhinagar, Ahmedabad, India}

\date{Recieved date; Accepted date}

\pacs{31.30.js;31.15.A-;31.15.vj;31.15.bw}

\vskip1.0cm

\begin{abstract}
\noindent
We investigate roles of electron correlation effects in the determination of the $g_j$ factors of the $4s ~ ^2S_{1/2}$, $4p ~ ^2P_{1/2}$, 
$4p ~ ^2P_{3/2}$, $3d ~ ^2D_{3/2}$, and $3d ~ ^2D_{5/2}$ states, representing to different parities and angular momenta, of the Ca$^+$ ion. 
Correlation contributions are highlighted with respect to the mean-field values evaluated using the Dirac-Hartree-Fock method, relativistic 
second order many-body theory, and relativistic coupled-cluster (RCC) theory with the singles and doubles approximation considering only the 
linear terms and also accounting for all the non-linear terms. This shows that it is difficult to achieve reasonably accurate results employing 
an approximated perturbative approach. We also find that contributions through the non-linear terms and higher-level excitations such as triple 
excitations, estimated perturbatively in the RCC method, are found to be crucial to attain precise values of the $g_j$ factors in the considered 
states of Ca$^+$ ion.  
\end{abstract}

\maketitle

\section{Introduction}

Spectroscopic studies of the singly ionized calcium (Ca$^+$) ion is of immense interest to both the experimentalists and theoreticians 
on many scientific applications. Particularly, this ion is under consideration for a number of high precision experimental studies such as in 
the atomic clock \cite{Chwalla,Huang}, quantum computation \cite{Riebe,Harty,Ballance}, testing Lorentz symmetry violation  
\cite{Pruttivarasin}, etc. A number of theoretical investigations have also been carried out in the determination of different physical
quantities by employing varieties of many-body methods \cite{Tommaseo,Itano,bksahoo1,Safronova1,bksahoo2,Safronova2,Mitroy,bksahoo3}, which 
demonstrate successfully achieving most of these properties meticulously compared to the experimental results. 

On the otherhand, there have been attempts to determine Lande $g_j$ factors in the atomic systems to ultra-high accuracy \cite{Tommaseo,Lindroth,
Shabaev1,Shabaev2}. The main motivation of these studies was to test validity of both the theories and measurements. Mostly, atomic systems with few 
electrons are being considered in these investigations aiming to find out role of higher order quantum electro-dynamics (QED) effects \cite{Shabaev2}. 
In these systems, both the QED and electron correlation effects contribute at par to match the theoretical calculations with the experimental 
results. Comparatively, only few attempts have been made to reproduce the experimental values of the $g_j$ factors to very high precision in the 
many-electron systems \cite{Tommaseo,Lindroth}. In the neutral or singly ionized heavy atomic systems, the electron correlation effects play the 
dominant roles for estimating the $g_j$ factors of the atomic states accurately. However, none of the previous studies have demonstrated the roles of electron 
correlation effects explicitly arising through various physical effects in the determination of the total values of the $g_j$ factors of the 
heavy atomic systems. Lindroth and Ynnerman had carried out such a rigorous investigation on the role of electron correlation effects to the 
corrections over Dirac value of the $g_j$ factors of the ground states in the Li, Be$^+$ and Ba$^+$ atomic systems, which have a valence electron in the $s$ 
orbital. They had employed a relativistic coupled-cluster (RCC) method and incorporated Breit interaction in their calculations and found that 
higher order correlation effects and Breit interaction play significant roles in achieving precise results. However, they had observed that lower
order contributions are still dominant in the evaluation of the corrections over Dirac value of the $g_j$ factors.  Especially, they had observed
that correlations due to all order core-polarization effects, arising through random-phase approximation (RPA) type of diagrams, in these 
calculations are crucial. A number of calculations have reported very accurate values of this quantity using multi-configuration Dirac-Fock 
(MCDF) method highlighting the importance of including the higher excited configuration state functions (CSFs) for their determinations 
\cite{Tommaseo,Cheng}. Shortcoming of this method is that it cannot explain roles of different electron correlation effects explicitly except 
giving a qualitative idea on their importance for incorporating to achieve precise results. Other points to be noted is that the MCDF method is a 
special case of the configuration interaction (CI) method. It is known that a truncated CI method have size consistent and size extensivity 
problems \cite{Bartlett,Szabo}. Moreover in practice, only the important contributing CSFs are being selected in this approach till the final results are 
achieved within the intended accuracies. In contrast, the truncated many-body methods formulated in the RCC theory framework are more capable of 
capturing the electron correlation effects rigorously than other existing atomic many-body methods and are also free from the size extensivity and
size consistent problems owing to exponential ansatz of the wave functions \cite{Bartlett,Szabo}. This is why RCC methods are generally termed as 
the golden tools for investigating roles of electron correlation effects in the spectroscopic studies. A number of properties in Ca$^+$ have been
calculated employing the RCC methods in the singles and doubles approximation (CCSD method) \cite{bksahoo1,Safronova1,bksahoo2,Safronova2,csur,
bksahoo4}. From these studies, the CCSD method and its equivalent level of approximated RCC methods are proven to be capable of giving very 
accurate results in the atomic systems having similar configurations with Ca$^+$. Thus, it would be interesting to learn how differently electron 
effects behave in the evaluation of the total values of the $g_j$ factors of the ground state as well as of the excited states belonging to both 
the parities and higher orbital angular momenta of an alkali-like atomic system like Ca$^+$. The present work is intended to demonstrate this by 
carrying out calculations of the $g_j$ factors of the $4s ~ ^2S_{1/2}$, $4p ~ ^2P_{1/2}$, $4p ~ ^2P_{3/2}$, $3d ~ ^2D_{3/2}$, and $3d ^2D_{5/2}$ 
states in the Ca$^+$ ion. 

\section{Theory}

The interaction Hamiltonian of an atomic electron when subjected to an external homogeneous magnetic field ${\vec {\bf B}}$ 
is given by \cite{Sakurai}
\begin{eqnarray}
 H_{mag} &=& ec \sum_i \mbox{\boldmath${\vec \alpha}$}_i \cdot {\vec \textbf{A}}_i \nonumber \\
         &=& - \frac{ec}{2} \sum_i \mbox{\boldmath${\vec \alpha}$}_i \cdot ({\vec \textbf{r}}_i \times {\vec \textbf{B}} ),
\end{eqnarray}
where $e$ is the electric charge of the electron, $c$ is the speed of light, $\mbox{\boldmath${\vec \alpha}$}$ is the Dirac operator, and ${\vec \textbf{A}}$ is 
the vector field seen by the electron located at $r$ due to the applied magnetic field. This interaction Hamiltonian can be expressed  
in terms of a scalar product as 
\begin{eqnarray}
 H_{mag} &=& -\frac{ec}{2} \sum_i ( \mbox{\boldmath${\vec \alpha}$}_i \times {\vec \textbf{r}}_i) \cdot {\vec \textbf{B}}  \nonumber \\
 &=& i \frac{ec}{\sqrt{2}} \sum_i r_i \left \{ \mbox{\boldmath${\vec \alpha}$}_i \otimes {\vec \textbf{C}}^{(1)} \right \}^{(1)} \cdot {\vec {\textbf B}} 
 \end{eqnarray}
with $C^{(1)}$ is the Racah coefficient of rank one.

Defining the above expression as $H_{mag} = {\bf \vec {\cal M}} \cdot {\vec {\textbf B}}$ with magnetic moment operator 
${\vec {\cal M}} = \sum_{i, q=-1,0,1} \mbox{\boldmath${\vec \mu}$}_q^{(1)} (r_i)$, the Dirac value to the Lande $g_j$ factor of a bound 
electron in an atomic system can be given by
\begin{eqnarray}
g_j^D = -\frac{1}{\mu_B} \frac{{\bf \vec {\cal M}}}{ \vec \textbf{J}} 
\end{eqnarray}
of a state with total angular momentum $\textbf{J}$ for the Bohr magneton $\mu_B = e \hbar / 2m_e$ with mass of electron $m_e$. Thus,
the $g_j^D$ value for the state $\vert J M \rangle$ can be evaluated using the projection theorem as
\begin{eqnarray}
g_j^D= - \frac{1}{2 \mu_B } \frac{\langle J|| {\bf {\cal M}} ||J \rangle}{ \sqrt{J(J+1)(2J+1)}},
 \label{eqn5}
\end{eqnarray}
with the corresponding single particle reduced matrix element of $\mbox{\boldmath${\mu}$}^{(1)}$ given by
\begin{eqnarray} 
\langle \kappa_{f}||\mbox{\boldmath${ \mu}$}^{(1)}||\kappa_{i}\rangle &=&-(\kappa_{f}+\kappa_{i})\langle -\kappa_{f}||\textbf{C}^{(1)}||\kappa_{i}\rangle \nonumber \\
&&\times\int^{\infty}_{0}dr \ r \ \left (P_{f}Q_{i}+Q_{f}P_{i} \right) ,
\label{eqn6}
\end{eqnarray}
where $P(r)$ and $Q(r)$ denote for the large and small components of the radial parts of the single particle Dirac orbitals, respectively, 
and $\kappa$ are their relativistic angular momentum quantum numbers. It can be noted here that this expression is similar to the expression 
for determining the magnetic dipole hyperfine structure constant, in both the properties the angular momentum selection rule is restricted by
the reduced matrix element of $C^{(1)}$, which is given as
\begin{eqnarray}
\langle \kappa_f\, ||\, \textbf{C}^{(k)}\,||\, \kappa_i \rangle &=& (-1)^{j_f+1/2} \sqrt{(2j_f+1)(2j_i+1)} \ \ \ \ \ \ \ \ \nonumber \\
                  &&  \left ( \begin{matrix} {
                          j_f & k & j_i \cr
                          1/2 & 0 & -1/2 \cr }
                         \end{matrix}
                            \right ) \Pi(l_{\kappa_f},k,l_{\kappa_i}), \ \ \ \ \
\end{eqnarray}
with
\begin{eqnarray}
\Pi(l_{\kappa_f},k,l_{\kappa_i}) &=&
\left\{\begin{array}{ll}
\displaystyle
1 & \mbox{for } l_{\kappa_f}+k+l_{\kappa_i}= \mbox{even}
\\ [2ex]
\displaystyle
0 & \mbox{otherwise,}
\end{array}\right.
\label{eqn12}
\end{eqnarray}
for the orbital momentum $l_{\kappa}$ of the corresponding orbital having the relativistic quantum number $\kappa$.

The net Lande $g$ factor of a free electron ($g_f$) with the QED correction on the Dirac value ($g_D$) can be approximately 
evaluated by \cite{Czarnecki}
\begin{eqnarray}
 g_f & \simeq & g_D \times \left [ 1 + \frac{1}{2} \frac{\alpha_e}{\pi } - 0.328 \left ( \frac{\alpha_e}{\pi} \right )^2 + \cdots  \right ] \nonumber \\
   &\approx & 1.001160 \times g_D ,
 \end{eqnarray}
where $\alpha_e$ is the fine structure constant. From this analysis, the QED correction to the bound electron $g_j$ factor can be 
estimated approximately by the interaction Hamiltonian as \cite{Akhiezer}
\begin{eqnarray}
 \Delta H_{mag} & \approx & 0.001160 \mu_B \beta \mbox{\boldmath${ \vec \Sigma}$} \cdot {\vec \textbf{B}}, 
\end{eqnarray}
where $\beta$ and $\mbox{\boldmath${ \vec \Sigma}$}$ are the Dirac matrix and spinor, respectively. Following the above procedure, we
can estimate leading order QED correction to $g_j$ by defining an operator ${\Delta \vec {\cal M}} = \sum_{i, q=-1,0,1} \Delta 
\mbox{\boldmath${\vec \mu}$}_q^{(1)} (r_i)=\sum_i \beta_i \mbox{\boldmath${ \vec \Sigma}$}_i$ such as \cite{Cheng}
\begin{eqnarray}
\Delta g_j^Q =0.001160 \frac{\langle J|| \Delta {\bf {\cal M}} ||J \rangle}{\sqrt{J(J+1)(2J+1)}}.
 \label{eqn7}
\end{eqnarray}
The corresponding reduced matrix element of the  $\Delta \mu^{(1)}_{q}(r_i)$ is given by
\begin{eqnarray}
\langle \kappa_{f}||\Delta \mbox{\boldmath${ \mu}$}^{(1)} || \kappa_{i}\rangle &=& (\kappa_f + \kappa_i -1)
  \langle - \kappa_{f}||\textbf{C}^{(1)}||\kappa_{i} \rangle \nonumber \\ && \times \int^{\infty}_{0}dr 
(P_{f}P_{i}+Q_{f}Q_{i}).
\label{eqn8}
\end{eqnarray}
 
 Hence, the total $g_j$ value of an atomic state can be evaluated as $g_j = g_j^D + \Delta g_j^Q$ and can be compared with the experimental 
value wherever available.

\section{Methods for Calculations}

 The considered states of Ca$^+$ have a common closed-core $[3p^6]$ of Ca$^{2+}$ with a valence orbital from different orbital angular momenta and 
parity. We have developed a number of relativistic many-body methods and have been employing them to calculate wave functions of a variety of atomic
systems including in Ca$^+$ that have configurations as a closed-core and a valence orbital \cite{bksahoo1,bksahoo2,csur,bksahoo4,Nandy,bksahoo5}.
Applications of these methods have proved that they are capable of giving rise very accurate results comparable with the experimental values. We 
apply some of these methods considering various levels of approximations to demonstrate how these methods are incapable of producing precise values 
of the $g_j$ factors in Ca$^+$. To find the reason for the same, the role of correlation effects at the lower and higher order contributions are 
investigated systematically. Special efforts have been made to estimate contributions from the leading triply excited configurations in the RCC 
theory framework adopting perturbative approaches to reduce the computational resources. For this purpose, we briefly discuss here the considered 
many-body methods and present results employing these methods to justify our above assessment.

To demonstrate various relativistic contributions systematically, we first perform calculations with the Dirac-Coulomb (DC) interaction and suppressing
contributions from the negative orbitals. In this approximation, the atomic Hamiltonian is given by 
\begin{eqnarray}
H^{\text{DC}} &=& \sum_i \Lambda_i^+ \left [ c\mbox{\boldmath$\vec \alpha$}_i\cdot {\vec \textbf{p}}_i+(\beta_i -1)c^2 + V_{\cal N}(r_i) \right ] \Lambda_i^+ \nonumber \\
  && + \sum_{i,j>i} \Lambda_i^+ \Lambda_j^+ \frac{1}{r_{ij}} \Lambda_i^+ \Lambda_j^+ , 
\end{eqnarray}
where $V_{\cal N}(r)$ is the nuclear potential and determined using the Fermi-charge distribution, $r_{ij}=|{\vec {\bf r}}_i-{\vec {\bf r}}_j|$
represents inter-electronic distance between the electrons located at $i$ and $j$, and $\Lambda^+$ operator represents a projection operator on to 
the positive energy orbitals. It is worth mentioning here is that the negative energy orbitals may contribute to quite significant, but it would be 
below the precision levels where the neglected electron correlation effects can also play dominant roles. That is the reason why we have not put
efforts to account for these contributions in the present work.

It is found in the previous calculation for the ground state of Ca$^+$, the frequency independent Breit interaction contributes sizably for the 
evaluation of the $g_j$ factor \cite{Tommaseo}. We also estimate contributions due to this interaction by adding the corresponding interaction 
potential energy expression in the atomic Hamiltonian as given by
\begin{eqnarray}
V_B(r_{ij})=-\frac{\{\mbox{\boldmath$\vec \alpha$}_i\cdot \mbox{\boldmath$ \vec \alpha$}_j+
(\mbox{\boldmath$\vec \alpha$}_i \cdot {\bf \hat{r}}_{ij})(\mbox{\boldmath$\vec \alpha$}_j\cdot {\bf \hat{r}}_{ij}) \}}{2r_{ij}} ,
\end{eqnarray}
where ${\bf \hat{r}}_{ij}$ is the unit vector along ${\bf \vec r}_{ij}$.

Apart from estimating $\Delta g_j^Q$ corrections to the $g_j$ factors due to the QED effects, it can be expected that there would be corrections 
to the $g_j^D$ values of the bound electrons from the modifications of the wave functions due to the QED effects. To estimate these corrections, 
we consider the lowest order QED interactions due to the vacuum potential (VP) and self-energy (SE) effects in the calculations of the wave 
functions of the bound electrons. The VP potential is considered as sum of the Uehling ($V_{U}(r)$) and Wichmann-Kroll ($V_{WK}(r)$) potentials, 
while the SE potential energy is evaluated as sum of the contributions from the electric and magnetic form-factors as were originally described 
in Ref. \cite{Flambaum}. The considered expressions with the Fermi charge distribution are given explicitly in our previous work \cite{bksahoo5}.

We first calculate the Dirac-Hartree-Fock (DHF) wave function of the $[3p^6]$ configuration ($\vert \Phi_0 \rangle$) using the above interactions 
in the atomic Hamiltonian. Then, the DHF wave function of a state of Ca$^+$ is constructed as $\vert \Phi_v \rangle= a_v^{\dagger} \vert \Phi_0 
\rangle$ with the respective valence orbital $v$ of the state. To show higher relativistic contributions explicitly, we perform calculations 
considering the DC Hamiltonian, then including the Breit interaction with the DC Hamiltonian, then with QED corrections in the DC Hamiltonian and 
finally, incorporating both the Breit and QED interactions simultaneously with the DC Hamiltonian. The reason for carrying out calculations 
considering individual relativistic corrections separately and then including them together is that we had observed in our previous study as sometimes 
correlations among the Breit and QED interactions alter the results than when they are incorporated independently. 

To investigate importance of electron correlation effects, we include them both in the lower order and all order many-body methods. In the lower 
order approximations, we employ the relativistic second order many-body perturbation theory (MBPT(2) method) and third order many-body perturbation 
theory (MBPT(3) method). In these approximations, we express the approximated atomic wave function as
\begin{eqnarray}
 \vert \Psi_v \rangle = \big{(} 1 + \Omega_0^{(1)} +\Omega_v^{(1)} \big{)} \vert \Phi_v \rangle,
\end{eqnarray}
in the MBPT(2) method and
\begin{eqnarray}
 \vert \Psi_v \rangle &=& \big{(} 1 + \Omega_0^{(1)} +\Omega_v^{(1)} + \Omega_0^{(1)} \Omega_v^{(1)} 
  +  \Omega_0^{(2)} +\Omega_v^{(2)} \big{)} \vert \Phi_v \rangle, \nonumber \\
\end{eqnarray}
in the MBPT(3) method, where $\Omega_0$ and $\Omega_v$ are known as wave operators. Here $\Omega_0$ and $\Omega_v$ act over $\vert \Phi_0 \rangle$
and $\vert \Phi_v \rangle$, respectively, to generate various CSFs in the perturbative approach. Amplitudes of these operators are determined by 
using the generalized Bloch's equations \cite{Lindgren} as 
\begin{eqnarray}
\langle \Phi_0^* \vert [\Omega_0^{(k)},H_0] \vert \Phi_0 \rangle &=& \langle \Phi_0^* \vert V_{es}(1+ \Omega_0^{(k-1)} ) \vert \Phi_0 \rangle 
\end{eqnarray}
and
\begin{eqnarray}
\langle \Phi_v^* \vert [\Omega_v^{(k)},H_0] \vert \Phi_v \rangle &=& \langle \Phi_v^* \vert V_{es} (1+ \Omega_0^{(k-1)}+ \Omega_v^{(k-1)}) \vert \Phi_v \rangle \nonumber \\
 & - & \sum_{m=1 }^{k-1} \langle \Phi_v^* \vert \Omega_v^{(k-m)} \vert \Phi_v \rangle  E_v^{(m)} , 
 \label{mbsv}
\end{eqnarray}
where $H_0$ is the DHF Hamiltonian, $V_{es}=H-H_0$ is the residual potential, $\vert \Phi_0^* \rangle $ and $\vert \Phi_v^* \rangle $ are the 
excited configurations over the respective $\vert \Phi_0 \rangle $ and $\vert \Phi_v \rangle $ DHF wave functions, and $E_v^{(k)} = 
\langle \Phi_v \vert V_{es} (1+ \Omega_0^{(k-1)}+\Omega_v^{(k-1)}) \vert \Phi_v \rangle$ is the $k^{th}$ order energy of the $\vert \Psi_v \rangle$ 
state.

\begin{table}[t]
\caption{Electron attachment energies (in cm$^{-1}$) using relativistic many-body methods at different levels of approximations with 
the DC Hamiltonian. Higher order relativistic corrections from the Breit interaction and QED effects are quoted from the CCSD method 
considering them separately and including together (given as ``Breit$+$QED''). Our final CCSD results are compared with the experimental values (mentioned as ``Expt'') 
listed in the NIST database \cite{nist}.}
\begin{ruledtabular}
\begin{tabular}{lccccc} 
 Method  & $4s \ ^2S_{1/2}$ & $3d \ ^2D_{3/2}$ & $3d \ ^2D_{5/2}$ & $4p \ ^2P_{1/2}$ & $4p \ ^2P_{3/2}$ \\
 \hline
  & & \\
 DHF & 91439.97 & 72617.49 & 72593.39 & 68036.82 & 67837.16 \\
 MBPT(2) & 96542.41 & 83943.81 & 83372.99 & 71026.03 & 70654.06 \\
 LCCSD & 96737.80 & 84564.90 & 84397.55 & 71101.05 & 70862.78 \\
 CCSD & 95879.60 & 81695.19 & 81606.44 & 70603.50 & 70372.14 \\
 & & \\
 \multicolumn{6}{c}{Relativistic corrections } \\
  Breit & $-7.42$ & $37.98$ & $53.15$ & $-11.02$ & $-3.70$ \\
  QED   & $-5.68$ & $2.11$ & $2.52$ & $0.02$ & $0.66$ \\
  Breit$+$ & $-13.09$ & $40.08$ & $55.67$ & $-11.01$ & $-3.05$ \\
  QED  \\
  \hline 
       &          &          &           &           &          \\
 Total & 95866.51 & 81735.27 & 81662.11  &  70592.49 &  70369.09 \\
 Expt & 95751.87(3) & 82101.68 &  82040.99 & 70560.36 & 70337.47 \\
\end{tabular}
\end{ruledtabular}
\label{tab1}
\end{table}

\begin{table*}[t]
\caption{Demonstration of trends of the calculated $g_j^D$ values in various relativistic methods using the DC Hamiltonian. Relativistic 
corrections from the CCSD method and contributions from the important triple excitations are given separately. Contributions to $\Delta g_j^Q$ 
at the DHF and CCSD method are also listed with the DC Hamiltonian. Accounting both the $g_j^D$ and $\Delta g_j^Q$ values, the net $g_j$ values 
are estimated in the DHF and CCSD methods to compare them with the available experimental results in the $4s \ ^2S_{1/2}$ and $3d \ ^2D_{5/2}$ 
states.}
\begin{ruledtabular}
\begin{tabular}{lccccc} 
    & $4s \ ^2S_{1/2}$ & $3d \ ^2D_{3/2}$ & $3d \ ^2D_{5/2}$ & $4p \ ^2P_{1/2}$ & $4p \ ^2P_{3/2}$ \\
 \hline
  & & \\
  \multicolumn{6}{c}{DC contributions at the DHF method} \\
 $g_j^D$ & 1.999953 & 0.799922 & 1.199917 & 0.666636 & 1.333308 \\
 $\Delta g_j^Q$ & 0.002320 & $-0.000464$  & $0.000464$  & $-0.000773$ & 0.000773 \\
 & & \\
 Net $g_j$ & 2.002273 & 0.799458 & 1.200381 & 0.665863 & 1.334081 \\
 \hline 
 & & \\
  \multicolumn{6}{c}{DC contributions to $g_j^D$} \\
 MBPT(2) & 1.999551 & 0.798641 & 1.197217 & 0.666457 & 1.333004 \\
 MBPT(3) & 1.999997 & 0.782330 & 1.186208 & 0.669819 & 1.333903 \\
 LCCSD & 1.996755 & 0.800997 & 1.197612 & 0.666674 & 1.332832 \\
 & & \\
  \multicolumn{6}{c}{ CCSD results to $g_j^D$} \\
 DC & 2.000654 & 0.799512 & 1.200430 & 0.666685 & 1.333521 \\
  $+$Breit & 2.000651 & 0.799550 & 1.200438 & 0.666689 & 1.333522 \\
  $+$QED   & 2.000651 & 0.799550 & 1.200438 & 0.666689 & 1.333522 \\
  $+$Breit$+$QED & 2.000650 & 0.799550 & 1.200438 & 0.666690 & 1.333522 \\
  $+$Triples & 1.999946 & 0.799019 & 1.199876 & 0.666409  & 1.333088 \\
& & \\
  $\Delta g_j^Q$ & 0.002321 & $-0.000465$  & $0.000465$  & $-0.000773$ & 0.000773 \\
  Net $g_j$ & 2.002267 & 0.79855 & 1.200341 & 0.665636 & 1.333861 \\
  \hline 
  & & \\
 Experiment & 2.00225664(9) \cite{Tommaseo} &  & 1.2003340(25) \cite{Chwalla} & & \\
\end{tabular}
\end{ruledtabular}
\label{tab2}
\end{table*}

 After obtaining amplitudes of the MBPT operators, the $g_j$ factors are calculated using the expression
\begin{eqnarray}
\langle O \rangle = \frac{\langle \Psi_v \vert O \vert \Psi_v \rangle}{ \langle \Psi_v \vert \Psi_v\rangle } ,
\label{preq}
\end{eqnarray}
where $O$ stands for the respective ${\bf {\cal M}}$ and $\Delta {\bf {\cal M}}$ operators for the evaluations of the $g_j^D$ and $\Delta g_j^Q$
contributions.

In the similar framework and using the exponential ansatz of RCC theory, atomic wave functions of the considered states with the respective 
valence orbitals are expressed as  
\begin{eqnarray}
 \vert \Psi_v \rangle = e^T \{ 1+ S_v \} \vert \Phi_v \rangle ,
 \label{eqcc}
\end{eqnarray}
where $T$ and $S_v$ are the RCC operators that excite electrons from $\vert \Phi_0 \rangle $ and $\vert \Phi_v \rangle $, respectively. We have 
approximated RCC theory to only the singles and doubles excitations (CCSD method). The single and double excitation processes carried out by these 
RCC operators are described by denoting these operators using the subscripts $1$ and $2$, respectively, as
\begin{eqnarray}
 T \simeq T_1 +T_2 \ \ \ \text{and} \ \ \ S_v \simeq S_{1v} + S_{2v}.
\end{eqnarray}
The amplitudes of these operators are evaluated by solving the equations
\begin{eqnarray}
 \langle \Phi_0^* \vert \overline{H}_N  \vert \Phi_0 \rangle &=& 0
\label{eqt}
 \end{eqnarray}
and 
\begin{eqnarray}
 \langle \Phi_v^* \vert \big ( \overline{H}_N - \Delta E_v \big ) S_v \vert \Phi_v \rangle &=&  - \langle \Phi_v^* \vert \overline{H}_N \vert \Phi_v \rangle ,
\label{eqsv}
 \end{eqnarray}
where $\vert \Phi_0^* \rangle$ and $\vert \Phi_v^* \rangle$ are excited up to doubles,  $\overline{H}_N= \big ( H_N e^T \big )_l$ represents 
for the linked terms only with the normal order Hamiltonian $H_N= H - \langle \Phi_0 \vert H \vert \Phi_0 \rangle$ and $\Delta E_v$ is the attachment 
energy for the state $\vert \Psi_v \rangle$, which is determined by
\begin{eqnarray}
 \Delta E_v  = \langle \Phi_v \vert \overline{H}_N \left \{ 1+S_v \right \} \vert \Phi_v \rangle .
 \label{eqeng}
\end{eqnarray}

To investigate the roles of the electron correlation effects through the non-linear terms in the RCC theory, we also perform calculations considering 
only linear terms in the singles and doubles approximation in this theory (which is termed as LCCSD method). In this approximation, it yields
\begin{eqnarray}
 \vert \Psi_v \rangle & \approx & \{ 1+ T + S_v \} \vert \Phi_v \rangle , \\
  \overline{H}_N & \approx & H_N + H_N T 
 \label{eqlcc}
\end{eqnarray}
and
\begin{eqnarray}
  \overline{H}_N S_v \approx H_N + H_N T + H_N S_v.
 \label{eqleng}
\end{eqnarray}

  After obtaining amplitudes of the RCC operators, the expectation values as in Eq. (\ref{preq}) are evaluated by
\begin{eqnarray}
\frac{\langle \Psi_v \vert O \vert \Psi_v \rangle}{ \langle \Psi_v \vert \Psi_v\rangle } 
&=& \frac{\langle \Phi_v | \{1+ T^{\dagger} + S_v^{\dagger}\} O \{1+ T+ S_v \} | \Phi_v \rangle} {\langle \Phi_v | \{1+ T^{\dagger} + S_v^{\dagger}\} 
\{1+ T + S_v \} | \Phi_v \rangle } \ \ \ \ 
\label{prpleq}
\end{eqnarray}
in the LCCSD method and
\begin{eqnarray}
\frac{\langle \Psi_v \vert O \vert \Psi_v \rangle}{ \langle \Psi_v \vert \Psi_v\rangle } 
&=& \frac{\langle \Phi_v | \{1+ S_v^{\dagger}\} e^{T^{\dagger}} O e^T \{1+S_v \} | \Phi_v \rangle} {\langle \Phi_v | \{1+ S_v^{\dagger}\} 
e^{T^{\dagger}} e^T \{1+S_v \} | \Phi_v \rangle } \ \ \ \ 
\label{prpeq}
\end{eqnarray}
in the CCSD method. Clearly, the expression for the LCCSD method gives rise finite number of terms like in the MBPT(2) and MBPT(3) methods.
However, the expression for the CCSD method has two non-terminating series in the numerator and denominator as $e^{T^{\dagger}} O e^T$ and 
$e^{T^{\dagger}} e^T$ respectively. These non-truncative series give a large number of non-linear terms corroborating a large space of CSFs 
belonging to higher level of excitations. To account for contributions from both the non-truncative series, we adopt iterative procedures. This is 
done by performing calculations through intermediate steps, in which we compute and store first the $O+OT+ T^{\dagger}O + T^{\dagger}O T$ terms 
from $e^{T^{\dagger}} O e^T$ and $1+ T^{\dagger}T$ terms from $e^{T^{\dagger}} e^T$. Then, we operate a $T$ operator and subsequently by a 
$T^{\dagger}$ operator on the above intermediate calculations and replace them as the new intermediate calculations. This procedure is 
repeated till we attain contributions up to 10$^{-8}$ precision level convergence in the values from the higher non-linear terms. 

\begin{figure}[t]
\begin{center}
\includegraphics[width=8.5cm,height=6.0cm]{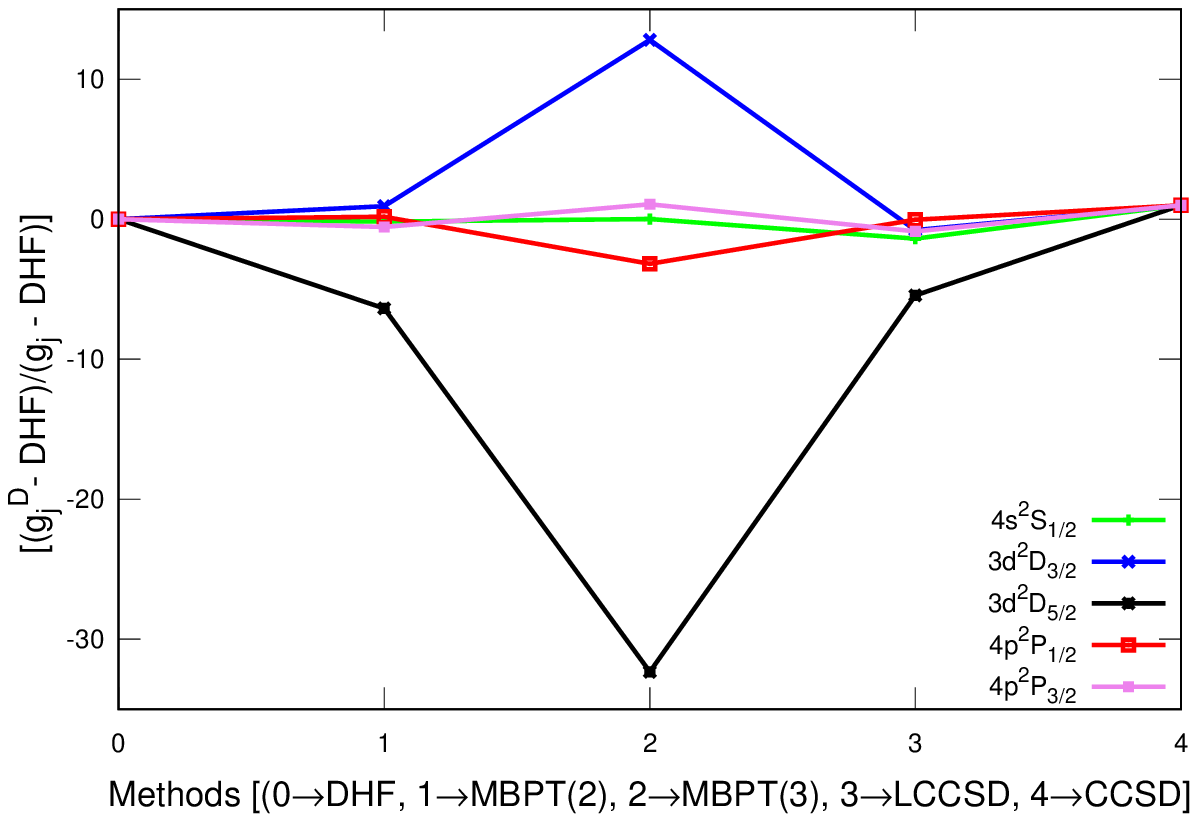}
\end{center}
\caption{Trends of electron correlation effects among the $4s \ ^2S_{1/2}$, $3d \ ^2D_{3/2}$, $3d \ ^2D_{5/2}$, $4p \ ^2P_{1/2}$ and $4p \ ^2P_{3/2}$ 
states for the evaluation of $g_j^D$ values in Ca$^+$. We plot the relative $(g_j^D-$DHF)/($g_j-$DHF) values to highlight the roles of correlation effects 
through different many-body methods. We consider values from the DHF, MBPT(2), MBPT(3), LCCSD and CCSD methods in a sequence referring them in the X-axis
in an arbitrary unit distance.}
\label{fig1}
\end{figure}

As we shall see, the correlation effects coming through the CCSD terms give much larger magnitudes to the $g_j$ factors than the available 
experimental values for the ground \cite{Tommaseo} and $3d ~ ^2D_{5/2}$ \cite{Chwalla} states of Ca$^+$, even though this method was proven 
to give reasonably accurate results for a number of properties in the considered ion as stated in Introduction. To find out how the higher
level excitations would circumvent this to bring back the results close to the experimental values, we define RCC operators in a perturbative 
framework to account for contributions from the important triply excited configurations from both $\vert \Phi_0 \rangle $ and 
$\vert \Phi_v \rangle $ as 
\begin{eqnarray}
 T_{3}^{pert}  &=& \frac{1}{6} \sum_{abc,pqr} \frac{\big ( H_N T_2 \big )_{abc}^{pqr}}{\epsilon_a + \epsilon_b + \epsilon_c - \epsilon_p -\epsilon_q - \epsilon_r} 
\label{t3eq}
 \end{eqnarray}
 and
\begin{eqnarray}
 S_{3v}^{pert} &=& \frac{1}{4} \sum_{ab,pqr} \frac{\big ( H_N T_2 + H_N S_{2v} \big )_{abv}^{pqr}}{\Delta E_v + \epsilon_a + \epsilon_b - \epsilon_p -\epsilon_q - \epsilon_r} ,
\label{s3eq}
 \end{eqnarray}
 where $\{a,b,c \}$ and $\{ p,q,r \}$ represent for the occupied and virtual orbitals, respectively, and $\epsilon$s are their single particle orbital energies. 
Contributions from the $T_{3}^{pert}$ and $S_{3v}^{pert}$ operators to the $g_j$ factors are estimated using Eq. (\ref{prpeq}) considering them as part of the $T$
and $S_v$ operators. In this approach, we evaluate extra terms as $T_2^{\dagger}OT_{3}^{pert}$, $T_2^{\dagger}OS_{3v}^{pert}$, $S_{2v}^{\dagger}OS_{3v}^{pert}$, 
$S_{1v}^{\dagger}T_2^{\dagger}OS_{3v}^{pert}$, $T_{3}^{pert \dagger}OT_{3}^{pert}$, $S_{3v}^{pert \dagger}OS_{3v}^{pert}$, and their complex conjugate (c.c.) terms.
These terms are computationally very expensive and give more than 500 Goldstone diagrams, but found to be crucial in achieving reasonably accurate results compared 
to the available experimental values.

\section{Results and Discussion}

\begin{table*}[t]
\caption{Contributions from individual CCSD terms to the $g_j^D$ values in the $4s \ ^2S_{1/2}$, $3d \ ^2D_{3/2}$, $3d \ ^2D_{5/2}$, $4p \ ^2P_{1/2}$, and 
$4p \ ^2P_{3/2}$ states. Contributions quoted as ``Extra'' and ``Norm'' are obtained from the rest of the non-linear terms of the CCSD method 
that are not listed here and corrections due to normalization of the wave functions, respectively. Values up to only the sixth decimal place are given and those 
values are finite but contribute below $10^{-6}$ precision level are quoted as $\sim 0.0$. Unusually large contributions coming from the correlation effects are 
highlighted by quoting them in bold. We have also underlined the ``Extra'' contribution to the $3d \ ^2D_{5/2}$ state to draw its attention for 
its very large value.}
\begin{ruledtabular}
\begin{tabular}{lccccc}
RCC terms & $4s \ ^2S_{1/2}$ & $3d \ ^2D_{3/2}$ & $3d \ ^2D_{5/2}$ & $4p \ ^2P_{1/2}$ & $4p \ ^2P_{3/2}$ \\
\hline        
& & \\
$O$                            &$1.999953$ &$0.799922$ &$1.199917$&$0.666636$ &$1.333308$ \\
 $OT_1 +$c.c.                  &$\sim 0.0$ &$0.0$      &$0.0$     &$\sim 0.0$ &$\sim 0.0$ \\
\vspace{0.3mm}
$T_1^{\dagger} O T_1$          &$0.000001$ &$0.0$      &$0.0$     &$0.000003$ &$0.000007$  \\
\vspace{0.3mm}
$T_1^{\dagger} O T_2+$c.c.     &$\sim 0.0$  &$0.0$     &$0.0$     &$\sim 0.0$ &$\sim 0.0$  \\
\vspace{0.3mm}
$T_2^{\dagger} O T_2$          &$-0.000912$ &$-0.006104$&$-0.009066$&$-0.000234$&$-0.000525$\\
\vspace{0.3mm}
$OS_{1v}+$c.c.                 &$-0.000009$ &$-0.000016$&$-0.000018$&$-0.000003$&$-0.000005$\\
\vspace{0.3mm}
$OS_{2v}+$c.c.                 &$0.000001$  &$-0.000004$&$0.000005$ &$-0.000003$&$-0.000003$ \\
\vspace{0.3mm}
$T_1^{\dagger} OS_{2v}+$c.c.   &$-0.000984$ &$-0.001526$&$-0.002282$&$-0.000144$&$-0.000284$ \\
\vspace{0.3mm}
$T_2^{\dagger}OS_{2v}+$c.c.    &$\sim 0.0$  &$\sim 0.0$ &$\sim 0.0$ &$\sim 0.0$ &$\sim 0.0$  \\
\vspace{0.3mm}
$S_{1v}^{\dagger}OS_{1v}$      &${\bf 0.005060}$  &${\bf 0.009181}$ &${\bf 0.013637}$ &${\bf 0.001606}$ &${\bf 0.003169}$  \\
\vspace{0.3mm}
$S_{1v}^{\dagger}OS_{2v}$+c.c. &$\sim 0.0$  &$\sim 0.0$ &$0.000001$ &$\sim 0.0$ &$\sim 0.0$  \\
\vspace{0.3mm}
$S_{2v}^{\dagger}OS_{2v}$      &${\bf 0.016159}$  &${\bf 0.018597}$ &$0.000008$ &${\bf 0.004210}$ &${\bf 0.007462}$  \\
\vspace{0.3mm}
$T_{2}^{\dagger}OT_{3}^{pert}+$c.c.   &  $\sim 0.0$ &  $\sim 0.0$  & $\sim 0.0$ & $\sim 0.0$ & $\sim 0.0$ \\
\vspace{0.3mm}
$S_{2v}^{\dagger}OT_{3}^{pert}+$c.c.   &  $0.0$ &  $0.0$  & $0.0$ & $0.0$ & $0.0$ \\
\vspace{0.3mm}
$T_{2}^{\dagger}OS_{3v}^{pert}+$c.c.   & ${\bf -0.001650}$ & ${\bf -0.000907}$ & ${\bf -0.001104}$ & ${\bf -0.000440}$  &   ${\bf -0.000804}$ \\
\vspace{0.3mm}
$S_{2v}^{\dagger}OS_{3v}^{pert}+$c.c.   & $\sim 0.0$ & $\sim 0.0$ & $\sim 0.0$ & $\sim 0.0$ & $\sim 0.0$ \\
\vspace{0.3mm}
$T_{3}^{pert \dagger}OT_{3}^{pert}$   &  {\bf 0.000136} & {\bf 0.000159} &  {\bf 0.000235} & 0.000050  &  0.000098 \\
\vspace{0.3mm}
$S_{3v}^{pert \dagger}OS_{3v}^{pert}$   &  ${\bf 0.000728}$ & {\bf 0.000238} & {\bf 0.000366} & {\bf 0.000130} & {\bf 0.000255}  \\
\vspace{0.3mm}
$S_{1v}^{\dagger}T_{2}^{\dagger}OS_{3v}^{pert}+$c.c.   & $-0.000076$ & $-0.000021$ &  $-0.000007$ & $-0.000021$  &  $-0.000038$ \\
\vspace{0.3mm}
Extra                     &${\bf 0.002951}$&${\bf -0.001207}$&$\underline{\bf 0.026959}$& ${\bf -0.000313}$&${\bf 0.000411}$ \\
\vspace{0.3mm}
Norm                      &$-0.021567$&$-0.019331$&$-0.028730$&$-0.005073$&$-0.010019$ \\
\end{tabular} 
\end{ruledtabular}
\label{tab3}
\end{table*}

In order to gauge correctness of the wave functions obtained by employing many-body methods at different levels of approximations, we first present
electron attachment energies to the considered states of Ca$^+$ in Table \ref{tab1} and compare them with the experimental values listed in the 
National Institute of Science and Technology (NIST) database \cite{nist}. We consider only the $4s \ ^2S_{1/2}$, $3d \ ^2D_{3/2}$, $3d \ ^2D_{5/2}$,
$4p \ ^2P_{1/2}$ and $4p \ ^2P_{3/2}$ states of Ca$^+$ as the representative states with different angular momentum and parity for our 
investigation. As can be seen from this table, the DHF results differ significantly from the experimental values while the MBPT(2) values are 
larger than the experimental results. The LCCSD method does not seem to improve the calculations and give even larger values than the MBPT(2) 
results. However, the CCSD method brings down these results close to the experimental values. Corrections from the Breit and QED interactions are 
given separately in the same table from the CCSD method. They are also estimated by including both these interactions simultaneously. In 
this case, we find sum of the individual corrections and simultaneous account of these corrections, quoted as Breit$+$QED in the above table, 
give almost the same contributions. In our earlier work on the Cs atom, we had found similar behavior for the attachment energies but trends were 
exhibiting differently in the evaluation of the transition properties \cite{bksahoo5}. Nevertheless, the higher order relativistic corrections are 
also removing slightly the discrepancies among the CCSD results and experimental values of the energies. It may be possible that the omitted 
contributions from the triple excitations improve the CCSD values further.

After understanding the role of the electron correlation effects in the evaluation of the energies, we present the calculated $g_j$ values of the 
$4s \ ^2S_{1/2}$, $3d \ ^2D_{3/2}$, $3d \ ^2D_{5/2}$, $4p \ ^2P_{1/2}$ and $4p \ ^2P_{3/2}$ states of Ca$^+$ in Table \ref{tab2} from a number of 
methods approximating at different levels. This also includes all the methods that were considered for evaluating energies along with the 
MBPT(3) method, which involves energies from the MBPT(2) method. To highlight how the correlation effects propagate in these methods, we present 
results systematically from lower to all order LCCSD and CCSD methods. We present both the $g_j^D$ and $\Delta g_j^Q$ results from the DHF method in 
the beginning to appraise beforehand about how much the electron correlation effects may render to the contributions for yielding results close to 
the experimental values in the measured states. In case, we are able to achieve results agreeing with the experimental values for some states then 
it may be possible to predict these values for other states using the employed many-body methods where measurements are not carried out. From the 
analysis of behavior of the correlation effects in the determination of the attachment energies, it was obvious to us that there were large differences 
between the calculations obtained using the DHF method and the experimental values. When we compare the net $g_j$ values of the ground and $3d \ ^2D_{5/2}$
states, after adding up the $g_j^D$ and $\Delta g_j^Q$ values, with the experimental results \cite{Chwalla,Tommaseo} quoted at the end of the above table, 
it gives an impression that the electron correlation effects may not play strong roles for attaining calculated values matching with the experimental results.
So it is natural to assume that employment of a lower order method can suffice the purpose. In the experimental paper on the ground state result, the authors 
have also presented theoretical results by carrying out a rigorous calculation employing the MCDF method \cite{Tommaseo}. It is demonstrated there that a very
large configurational space was required to attain results matching with their measured value. It was also highlighted in that work that the Breit interaction 
contribution was essential in achieving high precision theoretical result. 

\begin{figure}[t]
\begin{center}
\includegraphics[width=8.5cm,height=10.0cm]{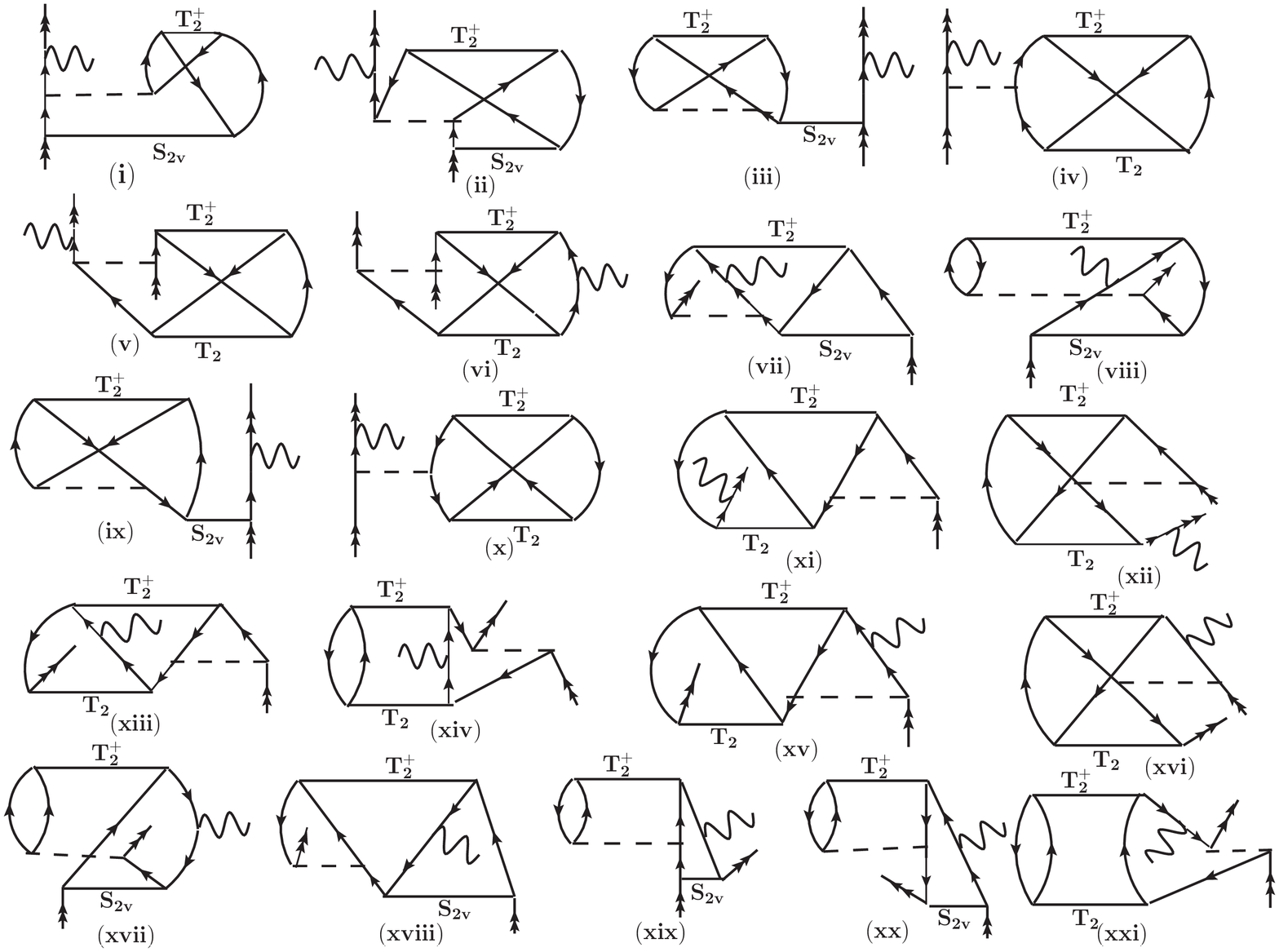}
\end{center}
\caption{Some of the important contributing Goldstone diagrams appearing through the $T_2^{\dagger} O S_{3v}^{pert}$ RCC term. Lines going up and down represent 
for the virtual and occupied orbitals of Ca$^+$. Lines with double arrows correspond to the valence orbital, dotted horizontal line means Coulomb interaction and 
solid horizontal lines correspond to the all order Coulomb interactions appearing through the $T_2$ and $S_{2v}$ operators.}
\label{fig2}
\end{figure}

As we move on, we shall explain the reasons why we shall not be able to achieve very high precision results by employing RCC theory in the CCSD 
method approximation. Thus, we do not prefer to present the calculated values of the $g_j$ factors beyond the sixth decimal places here. Necessity 
of including higher level excitations through the RCC method to improve these results further are demonstrated by investigating contributions from
the leading order triple excitation contributions involving the core and valence orbitals at the MBPT(3) method and in the perturbative approach 
using the RCC operators as defined in Eqs. (\ref{t3eq}) and (\ref{s3eq}). We have also quoted corrections to $g_j^D$ from the Breit and QED 
corrections considering them separately and also considering both the interactions together. The estimated $\Delta g_j^Q$ corrections from the 
CCSD method are also listed explicitly. Signs of these corrections are not the same for all the states owing to the $\kappa_f + \kappa_i -1$ factor 
in Eq. (\ref{eqn8}). It is obvious from Table \ref{tab2} that our CCSD results do not look very impressive when compared with the available
experimental values. However the important point to be noted from this work is on the trends of the results starting from the DHF method to the 
CCSD method, which shows how values are vacillating from one method to another in different states. 

 Since the differences among the values of the $g_j^D$ factors among various methods are very small, the role of electron correlation effects are 
not realized distinctly. To make it pronounced, we plot the ($g_j^D-$DHF)/($g_j-$DHF) values considering $g_j^D$ values from different methods in 
Fig. \ref{fig1} for all the states. It highlights the trends of the electron correlation effects incorporated through these methods. As can be seen 
from this figure, the correlation contributions do not follow definite trends and they are quite significant in view of achieving high precision 
values. Also, we give contributions to the $g_j^D$ values for all the considered states from the individual terms of the CCSD method including the 
terms including the perturbed triple excitations operators in Table \ref{tab3}. This is to notify how some of the higher order terms in the all order 
perturbative method contribute larger than the lower order RCC terms. The DHF value gives here the largest contribution as it includes the Dirac 
$g_D$ value. It has been found in the earlier studies on hyperfine structure constants and quadrupole moments of atomic states in $^{43}$Ca$^+$ 
using the RCC method \cite{bksahoo2} that after the DHF value, the dominant contributions come from the $OS_{1v}$ and $OS_{2v}$ terms along with 
their c.c. terms due to the electron correlation effects. It to be kept in mind that the $OS_{1v}$ term accounts for the lowest order electron 
pair-correlation effects, while the $OS_{2v}$ term incorporates the lowest order electron core-polarization effects in the RCC framework 
\cite{bksahoo6,bksahoo7}. The other terms encompass higher order correlation effects due to non-linear in RCC operators. Hence, it is generally 
anticipated that contributions from these non-linear terms are relatively smaller compared to the above two terms. However, we find in this case 
that many of the non-linear terms are giving much larger contributions, almost by an order, than the lower order RCC terms. Significantly 
contributing correlation effects are quoted in bold in the above table. Those non-linear terms from the CCSD method, which are not listed in the 
above table, their total contributions are given as ``Extra''. It is obvious from the above table that these contributions are quite large, 
especially in the $3d \ ^2D_{5/2}$ state which has been underlined. This suggests the core correlation contributions appearing through the 
$T$ operators in the non-linear terms play active roles in the evaluation of the $g_j^D$ values. Thus, it testifies that consideration of a 
perturbative method would completely fail to estimate the $g_j$ factors accurately in an atomic system. We had also seen in Table \ref{tab2} 
that contributions from the estimated triple excitations through the perturbed RCC operators are the decisive factors to attain the results close 
to the available experimental values. Following the perturbative analysis, it can be perceived that the $T_{2}^{\dagger}OT_3^{pert}$, 
$S_{2v}^{\dagger}OT_3^{pert}$, $T_{2}^{\dagger}OT_{3}^{pert}$ and $S_{2v}^{\dagger}OS_{3v}^{pert}$ RCC terms account for the lowest order 
terms involving the triply excited perturbed excitation operators. Since the $S_{3v}^{pert}$ operator involves the valence orbital, the term 
including this operator usually gives the larger contributions than the counter terms with the $T_{3}^{pert}$ operator. But comparison between 
the contributions obtained through the $T_{2}^{\dagger}OT_{3}^{pert}$,  $S_{2v}^{\dagger}OS_{3v}^{pert}$ and $S_{3v}^{pert \dagger}OS_{3v}^{pert}$ 
terms quoted in Table \ref{tab3} suggest that the correlation contributions do not manifest this trend. Analyzing in terms of level of excitations
associated with all these operators, as defined in Ref. \cite{Bartlett}, it can be understood that the Goldstone diagrams involving the 
particle-particle and hole-hole excitations through the ${\bf {\cal M}}$ operator are the important physical processes and the hole-particle 
and particle-hole excitations do not play much role in determining the $g_j^D$ values.

\begin{table*}[t]
\caption{Contributions to $g_j^D$ values of different states from the individual diagrams shown in Fig. \ref{fig2}. Values are given after multiplying with 
$10^{3}$ to highlight their contributions prominently and those values which are unusually large are quoted in bold. This clearly demonstrates importance of 
considering an all order perturbative method for the determination of the $g_j$ factors in the atomic systems.}
\begin{ruledtabular}
\begin{tabular}{lcccccccccc}
Diagrams & \multicolumn{2}{c}{$4s \ ^2S_{1/2}$} & \multicolumn{2}{c}{$3d \ ^2D_{3/2}$} & \multicolumn{2}{c}{$3d \ ^2D_{5/2}$} & \multicolumn{2}{c}{$4p \ ^2P_{1/2}$} & \multicolumn{2}{c}{$4p \ ^2P_{3/2}$} \\
\cline{2-3}       \cline{4-5} \cline{6-7} \cline{8-9} \cline{10-11}
& & \\
 &MBPT(3)& RCC& MBPT(3)& RCC& MBPT(3)& RCC& MBPT(3)& RCC& MBPT(3)& RCC \\
\cline{2-3}       \cline{4-5} \cline{6-7} \cline{8-9} \cline{10-11}
& & \\
Fig. \ref{fig2}(i)  & $0.4425$ & ${\bf 0.9141}$ & $0.3852$ & ${\bf 0.8401}$ & $0.5758$ & ${\bf 1.2575}$ & $0.0874$ & $0.1820$ & $0.1732$ &$0.3613$  \\
Fig. \ref{fig2}(ii) &$-0.1108$ &$-0.1652$ &$-0.1864$ &$-0.2811$&$-0.2800$&$-0.4228$&$-0.03306$&$-0.0565$ &$-0.0668$ &$-0.1143$  \\
Fig. \ref{fig2}(iii) &$0.2023$ &$0.3247$ & $0.4086$ &$0.6099$ &$0.6107$ &$0.9123$ &$0.0426$ &$0.0742$ &$0.0848$&$0.1477$ \\
Fig. \ref{fig2}(iv) & ${\bf 12.6380}$  &${\bf 20.5263}$ &${\bf 7.7036}$ &${\bf 12.2213}$ &${\bf 11.5434}$ &${\bf 18.3284}$&${\bf 3.5011}$&${\bf 5.5020}$&${\bf 6.9796}$&$10.9658$ \\
Fig. \ref{fig2}(v) &$-0.2017$ &$-0.2983$&$-0.5771$&$-0.9463$&$-0.8665$&$-1.4226$&$-0.03446$&$-0.0488$&$-0.0689$&$-0.9758$\\
Fig. \ref{fig2}(vi)  &$-0.2015$&$-0.2976$&$0.7874$&$1.3197$&$0.0436$ &$0.0882$&$0.0518$&$0.0766$&$-0.0256$&$-0.0344$ \\
Fig. \ref{fig2}(vii) &$-0.0913$  &$-0.1323$&$0.0596$ &$0.0845$&$-0.0843$&$-0.1372$&$\sim 0.0$&$\sim 0.0$&$-0.0385$&$-0.6697$ \\
Fig. \ref{fig2}(viii) &$0.2316$&$0.3907$&$-0.1859$&$-0.2984$&$0.0250$&$0.0662$&$\sim 0.0$&${\bf 0.0271}$&$0.0939$&$0.1601$ \\
Fig. \ref{fig2}(ix) &$-0.2204$&$-0.3402$ &$-0.4415$&$-0.6485$ &$-0.6600$&$-0.9701$&$-0.0459$&$-0.0770$&$-0.0915$&$-0.1533$  \\
Fig. \ref{fig2}(x) &${\bf -12.7009}$&${\bf -20.6074}$&${\bf -7.8297}$&${\bf -12.3756}$&${\bf -11.7321}$&${\bf -18.5594}$&${\bf -3.5141}$&${\bf -5.5182}$&${\bf -7.0057}$&${\bf -10.9981}$  \\
Fig. \ref{fig2}(xi) &$0.1408$  &$0.2022$& $0.5118$ &$0.8461$&$0.7691$&$1.2728$&$0.0270$&$0.0350$&$0.0544$&$0.0709$  \\
Fig. \ref{fig2}(xii) &$0.1322$&$0.1910$&$0.4986$&$0.8294$&$0.74897$&$1.2473$&$0.0256$&$0.0332$&$0.0516$&$0.0672$ \\
Fig. \ref{fig2}(xiii) &$0.1403$ &$0.2013$  &$-0.8225$&$-1.3819$&$-0.1216$ & $-0.2143$ & $-0.0442$&$-0.0620$&$0.0183$&$0.0213$\\
Fig. \ref{fig2}(xiv) &$-0.4488$&$-0.6347$ &$\sim 0.0$ & $0.0536$ &$-0.9322$&$-1.2610$&$-0.0238$&$-0.0225$&$-0.1680$&$-0.2248$ \\
Fig. \ref{fig2}(xv) &$0.1409$&$0.2023$&$0.5057$&$0.8341$&$0.7651$&$1.2648$&$1.2648$&$0.0315$&$0.05293$&$0.0692$ \\
Fig. \ref{fig2}(xvi) &$0.1323$ & $0.1911$ &$0.4847$ & $0.8071$ &$0.7399$&$1.2323$& $0.0202$&$0.0268$&$0.0489$&$0.0641$ \\
Fig. \ref{fig2}(xvii) &$-0.2309$&$-0.4032$&$0.1170$&$0.1908$&$-0.0707$&$-0.1466$&$0.0206$&$0.0371$&$-0.0793$&$-0.1334$\\
Fig. \ref{fig2}(xviii)   & $0.0914$ &$0.1369$ & $-0.0665$& $-0.1048$& $0.0801$&$ 0.1306$& $-0.0237$& $-0.0471$& $0.0294$&$0.0501$  \\
Fig. \ref{fig2}(xix)   & $0.0962$&$0.1688$ &$-0.0104$&$-0.0180$&$0.0496$&$0.0379$&$-0.0289$&$-0.0342$&$-0.0121$&$\sim 0.0$  \\
Fig. \ref{fig2}(xx)   &$\sim 0.0$&${\bf 0.2768}$&$\sim 0.0$&$\sim 0.0$&$0.8479$&$1.2160$&$\sim 0.0$& ${\bf -0.0553}$&$\sim 0.0$&${\bf 0.0784}$ \\
Fig. \ref{fig2}(xxi)   &$0.4482$&$0.6543$&$-0.1303$&$-0.1868$&${\bf 0.8479}$&$1.2160$&$\sim 0.0$&$\sim 0.0$&$0.1536$&$0.2175$\\
\end{tabular} 
\end{ruledtabular}
\label{tab4}
\end{table*}

 Again, we have observed that similar types of Goldstone diagrams attribute completely different trends of correlation effects at the lowest 
order and all order methods. To demonstrate it more prominently, we find out the leading order contributing diagrams from the $T_{2}^{\dagger}OT_{3}^{pert}$ and 
$S_{2v}^{\dagger}OS_{3v}^{pert}$ RCC terms and compare contributions from these diagrams with their counter lowest order Goldstone diagrams 
appearing through the MBPT(3) method. We have shown some of these diagrams in Fig. \ref{fig2} and quote their contributions in Table \ref{tab4}
from the MBPT(3) and RCC methods. As can be seen from this table, there are huge differences in some of the results obtained at the MBPT(3) 
method and at the level of RCC calculations. We have also quoted some contributions in bold to bring to the attention on the unusually large  
contributions at the lower and all order level calculations. Again, it is obvious from this table that some diagrams contribute predominantly 
to the lower angular momentum states while other diagrams contribute significantly in higher angular momentum states. Some changes in the 
correlation trends are also observed among the states belonging to different parities. 

Nonetheless, unusually large contributions arising through the perturbed triple excitation RCC operators implies that RCC theory in the 
CCSD method approximation is not capable of producing precise values of the $g_j$ factors in Ca$^+$. Also, larger contributions arising through
some of the non-linear terms than the linear terms in the CCSD method suggests that consideration of full triple excitations may be imperative  
to achieve $g_j$ factors below the $10^{-6}$ precision level. Moreover, either estimating the $g_j^D-g_D$ value as in Ref. \cite{Lindroth} or 
developments of alternative RCC theories, such as bi-orthogonal RCC theory \cite{Bartlett}, avoiding appearance of non-truncative series as in 
Eq. (\ref{prpeq}) to determine the $g_j$ factor of a state in this ion would be inevitable. 

\section{Conclusion}

We have employed a number of relativistic many-body methods to investigate roles of the electron correlation effects in the determination of the 
$g_j$ factors of the first five low-lying atomic states in the singly charged calcium ion. To validate these methods, we first present the electron
attachment energies by employing these methods and compare them against the experimental values listed in the National Institute of Science and 
Technology database. This demonstrates gradual improvement of accuracies in the results from lower many-body methods to all order relativistic 
coupled-cluster method with the singles and doubles approximation. However, when these methods are employed for the determination of the $g_j$ 
factors of the considered atomic states, the trends of the correlation effects were found to be very peculiar in nature. In fact, the results 
obtained employing the mean-field theory in the Dirac-Hartree-Fock approach are found to be in better agreement with the experimental values than 
the lower-order many-body perturbation theories and relativistic coupled-cluster theory with linear terms approximation. We also found that triple 
excitation contributions are the decisive factors in achieving very precise values for the $g_j$ factors and their contributions through the lower 
order and all order correlation effects behave completely different. Nonetheless, the overall observation from this study is that it is very 
challenging to attain high accuracy $g_j$ factors in many-electron systems by employing a truncated many-body method as the contributions from the 
electron correlation effects do not converge with the higher order approximations. Thus, it is reliable to determine the $g_j-g_D$ value instead 
of the net $g_j$ value of an atomic state. Also, it is imperative to develop more powerful relativistic many-body methods circumventing the 
problem of appearing non-truncative series so that trends of the correlation effects can be systematically investigated and calculations can be 
improved gradually in the determination of the $g_j$ factors in a many-electron atomic system. Since unique correlation effects are associated with 
the determination of $g_j$ factors, it suggests us that capable of a relativistic many-body method can be indeed scrutinized by producing high precision 
values for these factors in heavy atomic systems. This test would be of immense interest in a number of applications such as investigating parity 
non-conservation and frequency standard studies in atomic systems more reliably.

\section*{Acknowledgement}

Computations were carried out using the Vikram-100TF HPC cluster at the Physical Research Laboratory, Ahmedabad, India.

\end{document}